\documentclass[aps,prd,twocolumn,showpacs,amssymb]{revtex4-1}
\usepackage{graphicx}
\usepackage{bm}
\usepackage{amsmath}
\usepackage{amssymb}%
\newcommand{\apjs}{Astrophys. J. Suppl. S.}
\newcommand{\apjl}{Astrophys. J. Lett.}
\newcommand{\aap}{Astron. Astrophys.}
\newcommand{\physrep}{Phys. Rep.}
\newcommand{\apss}{Astrophys. Space Sci.}
\newcommand{\mnras}{Mon. Notices Royal Astron. Soc.}

\def\ba{\begin{eqnarray}}
\def\ea{\end{eqnarray}}

\begin{document}

\title{Evolution of inspiralling neutron star binaries: effects of tidal interactions and orbital eccentricities}
\author{Jie-Shuang Wang}
\email{jiesh.wang@gmail.com}
\affiliation{Tsung-Dao Lee Institute, Shanghai Jiao Tong University, Shanghai 200240, China}
\author{Dong Lai}
\affiliation{Cornell Center for Astrophysics and Planetary Science, Department of Astronomy, Cornell University, Ithaca, NY 14853, USA}
\affiliation{Tsung-Dao Lee Institute, Shanghai Jiao Tong University, Shanghai 200240, China}

\begin{abstract}
Neutron star (NS) binaries formed dynamically may have significant eccentricities while emitting gravitational waves (GWs) in the LIGO/VIRGO band. 
We study tidal effects in such eccentric inspiralling NS binaries using a set of hybrid equations. The NS is modelled as a compressible ellipsoid, which can deform nonlinearly due to tidal forces, while the orbit evolution is treated with the post-Newtonian (PN) theory up to 2.5-PN order. 
We find that in general, the tidal interaction can accelerate the inspiral, and cause orbital frequency and phase shifts. 
For circular inspirals, our calculations reproduce previous linear result at large binary separations, but incorporate the dynamical response of the NS at small separations. 
For eccentric inspirals, the frequency and phase shifts oscillate considerably near pericenter passages, and the oscillating amplitudes increase with eccentricities. 
As a result, the GW phase is also significantly influenced by the tidal effect. 
At merger, the cumulative GW phase shift can reach more than 10 radians (for typical NS mass $1.4M_\odot$ and radius 11.6 km), much larger than the circular inspiral case. 
Although the event rate of eccentric NS mergers is likely low, the detection of such mergers could provide a useful constraint on the NS equation of state. 
\end{abstract}
\maketitle

\section{Introduction}

Binary neutron star (BNS) mergers are one of the primary sources for multi-messenger astrophysics. They produce gravitational waves (GWs)
and copious electromagnetic signals. The recent LIGO detection of the
first binary neutron star merger event GW170817, together with
its joint detection in multiple electromagnetic bands, heralded the
beginning of multi-messenger astronomy \cite{Abbott2017}.

There are two broad classes of formation channels for merging double
compact object (CO) binary systems. The first is the standard isolated
binary evolution channel, where the binary components are brought
closer by dynamical frictions in mass-transferring or common envelope
phases \cite[e.g., see][and references therein]
{Flannery1975,Massevitch1976,Smarr1976,Belczynski2002,Kalogera2007,Postnov2014,Belczynski2018}.
CO binaries formed in this channel are expected to have circular
orbits while emitting high-frequency GWs in the LIGO/Virgo band. The
second class of formation channels involve gravitational interactions
between multiple stars and COs. For instance, black hole (BH) binaries in dense star clusters can become bound and shrink in separation due to
three-body encounters (e.g. an exchange interaction between a binary
and a CO) and/or secular interactions
\cite[e.g.][]{PZ2000,Miller2002,Wen2003,OLeary2009,Miller2009,Antonini2012,Rodriguez2015,Samsing2018}. Dynamical
formation of merging compact binaries can also occur in the galactic
field or near supermassive BHs, where CO mergers are induced
in hierarchical triple or quadruple systems
\cite{Silsbee2017,Antonini2017,Liu2018,Liu2019,Liu2019b}.
Eccentric inspirals of CO binaries in the LIGO/Virgo band
may be produced in these dynamical channels, although the formation rate is uncertain.

GWs from the final inspiral of BNS systems carry important
information of the matter properties under extreme conditions.
Tidal effects on the gravitational waveform in circular neutron star (NS) binary inspirals have been studied extensively \cite[e.g., see][and reference therein]{Kochanek1992,Bildsten1992,Lai1995,Lai1996,Baumgarte1998,Penner2011,Ferrari2012,Xu2017,Andersson2018},
including quasi-equilibrium tides
\cite{Lai1994a,Flanagan2008} and resonant tides
\cite{Lai1994b,Shibata1994,Reisenegger1994,Ho1999,Lai2006,Yu2017a,Yu2017b,Andersson2018,Xu2018,Yang2019}.
Observations of the NS merger events GW170817
\cite{LIGO2017} and GW190425 \cite{Ligo2020} provide direct
constraints on the dimensionless tidal deformability parameter, which
are then used to constrain the equation of state of the NS.

Recently, a number of studies have examined the tidal effects on
eccentric NS binary inspirals
\cite{Chirenti2017,Chaurasia2018,Yang2018,Yang2019,Vick2019}. These
works treat the tidal deformation or forced NS oscillation in the
linear approximation. This is expected to break down at small binary
separations, where the tidal effects on the gravitational waveform are
strongest. In this paper, we study the tidal effects on the orbital
evolution and gravitational waveform of eccentric NS binaries, going beyond the
linear approximation. We treat the NS as a deformable ellipsoid, and directly incorporate
the tidal force in the evolution of the orbit and the NS. The advantage of adopting
this ellipsoid model is that it includes the tidal effects in both the
linear and nonlinear regimes.
In Section \ref{models}, we introduce the dynamical equations used to
model the NS binary evolution. We present sample numerical results in
Section \ref{results}, and discuss the tidal effects on the GW signal in
Section \ref{GW}. The conclusion are presented in
Section \ref{Conclusion}.

\section{Dynamical equations of inspiraling NS binary\label{models}}

We consider an inspiraling NS binary as an example to explore the
tidal effects on the gravitational wave. We employ a set of hybrid
dynamical equations: The NS is modelled with the Newtonian dynamical
equations (Eqs.~\ref{eq:a1}-\ref{eq:lambda} below) in the compressible
ellipsoidal model, where the buoyancy effect is neglected; 
And we use the hybrid post-Newtonian equations of
motion (up to the 2.5-PN order) to evolve the orbital evolution
(Eqs.~\ref{eq:r} and \ref{eq:theta} below).

\begin{figure}[t]
  \centering
	\includegraphics[width=0.49\textwidth]{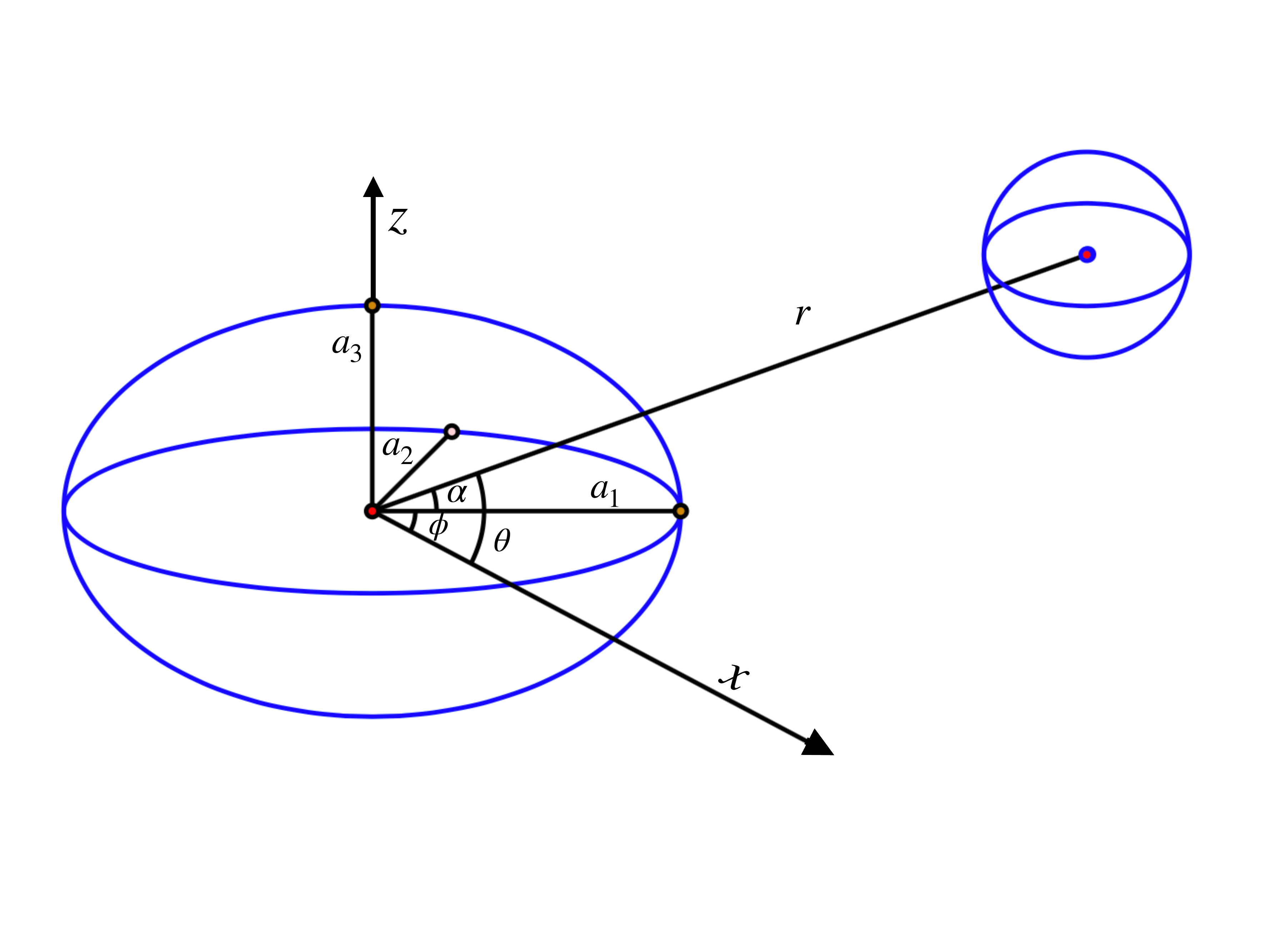}
	\caption{The coordinate system of a NS-BH binary. 
	 The $z$ axis is fixed to be parallel to the principal axis $a_3$ and is perpendicular to the
     orbital plane, while the $x$ and $y$ axes lie in the orbital plane and are
     parallel to $a_1$ and $a_2$ at $t=0$, respectively.}
	\label{fig:1}
\end{figure}

The NS is modelled as a compressible Riemann-S ellipsoid with mass
($M$) and a polytropic equation of state $P=K\rho^{1+1/n}$. The
ellipsoid is characterised by three principal axes ($a_1,~a_2$ and
$a_3$), the angular frequency of the ellipsoidal figure ($\Omega$),
and the angular frequency of the internal circulation ($\Lambda$)
\cite{Chandrasekhar1969,Lai1993}. The coordinate system is shown in
Fig. \ref{fig:1}, where we set $a_3$, $\Omega$, $\Lambda$ and the
orbital angular velocity vectors ($\Omega_{\rm orb}$) along the $z$-axis (perpendicular to the
orbital plane).
The NS orbits around a companion of mass $M'$ with separation $r(t)$
and orbital phase $\theta(t)$. 
The companion here is treated as a point mass and labelled as ``BH".
The dynamical equations for such a system can be 
expressed as follows \cite{Lai1994,Lai1995,Lai1996,Kidder1993}:
\begin{eqnarray}
\ddot{a}_1 &=& a_1(\Omega^2+\Lambda^2)-2a_2\Omega\Lambda -{2\pi\over q_n}a_1A_1\bar\rho \nonumber \\
& &+\left({5k_1 \over n\kappa_n}{P_c\over\rho_c}\right){1\over a_1}
+{M'a_1\over r^3}(3\cos^2\alpha-1),        \label{eq:a1}      \\
\ddot{a}_2 &=& a_2(\Omega^2+\Lambda^2)-2a_1\Omega\Lambda
-{2\pi\over q_n}a_2A_2\bar\rho \nonumber \\
& &+\left({5k_1 \over n\kappa_n}{P_c\over\rho_c}\right){1\over a_2}
+{M'a_2\over r^3}(3\sin^2\alpha-1),      \label{eq:a2}       \\
\ddot{a}_3 &=& -{2\pi\over q_n}a_3A_3\bar\rho
+\left({5k_1 \over n\kappa_n}{P_c\over\rho_c}\right){1\over a_3}
-{M'a_3\over r^3},              \label{eq:a3}        \\ 
\dot\Omega &=& \biggl[2\left({\Omega\over a_2}+{\Lambda\over a_1}\right)\dot a_1 
-2\left({\Omega\over a_1}+{\Lambda\over a_2}\right)\dot a_2 \nonumber \\
& &-{3M'\over 2r^3}\left({a_1\over a_2}+{a_2\over a_1}\right)\sin 2\alpha \biggr] 
\left({a_2\over a_1}-{a_1\over a_2}\right)^{-1}, \label{eq:omega}\\
\dot\Lambda &=& \biggl[2\left({\Omega\over a_1}+{\Lambda\over a_2}\right)\dot a_1
-2\left({\Omega\over a_2}+{\Lambda\over a_1}\right)\dot a_2 \nonumber \\
& &-{3M'\over r^3}\sin 2\alpha \biggr]\left({a_2\over a_1}-{a_1\over a_2}\right)^{-1}, \label{eq:lambda}	   \\
\ddot r &=& -{3\kappa_n\over 10}{M_t\over r^4}\left[a_1^2(3\cos^2\alpha-1)
+a_2^2(3\sin^2\alpha-1)-a_3^2\right] 
\nonumber \\
& &+r{\dot\theta}^2-{M_t\over r^2}(1+A_H+B_H\dot r+A_{5/2}+B_{5/2}\dot r)
, \label{eq:r}\\
\ddot\theta &=& -{M_t\over r^2}(B_H+B_{5/2})\dot\theta  -{3\kappa_n\over 10}{M_t\over r^5}(a_1^2-a_2^2)\sin 2\alpha \nonumber \\
& & -{2\dot r\dot\theta\over r}
.\label{eq:theta}
\end{eqnarray}
In the above, $M_t=M+M'$ is the total mass, $A_1,~A_2$ and $A_3$ are
dimensionless functions of $a_1,a_2,a_3$ as defined in Section 17 of
Ref. \cite{Chandrasekhar1969}. The constants $\kappa_n$,
$q_n\equiv\kappa_n(1-n/5)$ and $k_1$ depend only on $n$
\cite{Lai1993}. For $n\neq0$, the pressure term satisfies ${5k_1
 P_c/( n\kappa_n\rho_c)}=M/(q_nR_0)(R/R_0)^{-3/n}$
\cite{Lai1994,Lai1995}, where the mean radius is
$R\equiv(a_1a_2a_3)^{1/3}$ and $R_0$ is its initial value (with no
tidal deformation and rotation). The mean density is
$\bar\rho=3M/(4\pi a_1a_2a_3)$. The angle $\alpha=\theta-\phi$ is
related to the angular frequencies through
$\dot\theta\equiv\Omega_{\rm orb}$ and $\dot\phi\equiv\Omega$
($\Omega_{\rm orb}$ is the instantaneous orbital frequency and $\Omega$
is the rotation frequency of the ellipsoidal figure).
The various post-Newtonian correction terms are 
\begin{eqnarray}
A_H &=& -1+{1-M_t/r\over (1+M_t/r)^3}-\left[{2-M_t/r\over
1-(M_t/r)^2}\right]{M_t\over r}\dot r^2+v^2 \nonumber \\
& &-\eta\left(2{M_t\over r}-3v^2+{3\over 2}\dot r^2\right)
+\eta\biggl[{87\over 4}\left({M_t\over r}\right)^2 \nonumber\\
& &+(3-4\eta)v^4+{15\over 8}(1-3\eta)\dot r^4
-{3\over 2}(3-4\eta)v^2\dot r^2\nonumber\\
& &-{1\over 2}(13-4\eta){M_t\over r}v^2
-(25+2\eta){M_t\over r}\dot r^2\biggr] 
\\
B_H &=& -\left[{4-2M_t/r\over 1-(M_t/r)^2}\right]\dot r
+2\eta\dot r-{1\over 2}\eta\dot r\biggl[(15+4\eta)v^2\nonumber\\
& &-(41+8\eta){M_t\over r}-3(3+2\eta)\dot r^2\biggr], 
\\
A_{5/2} &=& -{8\over 5}\eta{M_t\over r}\dot r\left(18v^2
+{2\over 3}{M_t\over r}-25\dot r^2\right),
\\
B_{5/2} &=& {8\over 5}\eta{M_t\over r}\left(6v^2
-2{M_t\over r}-15\dot r^2\right) \; ,
\end{eqnarray}
where $v^2=\dot r^2+r^2\dot\theta^2$, 
$\eta=\mu/M_t$, $\mu=MM'/M_t$ and $I_{ii}=\kappa_nMa_i^2/5$ for $i=1,~2,~3$. 
Note that the PN terms in the orbital equations (\ref{eq:r})-(\ref{eq:theta}) are of the ``hybrid" form derived in Ref. \cite{Kidder1993}: These hybrid equations augment the Schwarzschild geodesic equations of motion with the finite-mass terms of the PN2.5 equations of motion; they properly account for the transition from orbital inspiral at large separations to plunge at small separations. 
This hybrid PN model is in essence similar to the effective one-body (EOB) model \cite{Buonanno1999}. We find it's convenient to incorporate the ellipsoid model in the hybrid PN formalism. A detailed comparison of our results with those using EOB models (e.g. Ref. \cite{Hinderer2016}) is beyond the scope of this paper.
As in previous semi-analytical works on eccentric mergers, our dynamical equations (\ref{eq:a1})-(\ref{eq:theta}) include the dominant gravitational radiation associated with the orbital motion, but not the enhanced radiation associated with the tidally deformed NS.
We estimate that this effect will contribute to about $\lesssim15\%$ to the phase error due to tidal effects (see the discussion below Eq. \ref{eq:dtheta_linear}).

\section{Numerical results\label{results}}

We consider a NS binary system with $M=M'=1.4M_\odot$,
$R_0=11.6$~km. The NS is modelled as a polytrope with $n=0.5$, as it represents a reasonable approximation to a large class of NS equations of state, especially stiff ones.
The initial condition is obtained by setting
$\ddot{a}_i=\dot{a}_i=\dot{\Omega}=\dot{\Lambda}=\ddot{r}=\dot{r}=\ddot{\theta}=0$
and $\alpha=\theta=\phi=0$ (so that $a_1$ and $a_2$ are along $x$ and
$y$ axes, respectively). The binary is initially at the apocenter
with separation $r_0=r_{p,0}(1+e_0)/(1-e_0)$, where $r_{p,0}$ is the initial pericenter distance. 
The initial eccentricity $e_0$ is defined in terms of the initial orbital angular
frequency $\Omega_{\rm orb,0}$ and $\Omega_{\rm circ,orb,0}$, the
angular frequency for circular orbit at $r_0$ (in the absence of tidal
effects and PN effects): $\sqrt{1-e_0}=\Omega_{\rm orb,0}/\Omega_{\rm
 circ,orb,0}$. We set the initial $\Omega_0=\Omega_{\rm orb,0}$, and
$\Lambda_0=2a_{1,0}a_{2,0}\Omega_0/(a_{1,0}^2+a_{2,0}^2)$,
corresponding to an {\it irrotational} NS \cite{Lai1993}).
We integrate Eqs.~(\ref{eq:a1})-(\ref{eq:theta}) forward in time
and stop the integration at $t=t_{\rm merge}$, when the binary separation decays to
$r=2.5R_0$.

\begin{figure}
\centering \includegraphics[width=0.5\textwidth]{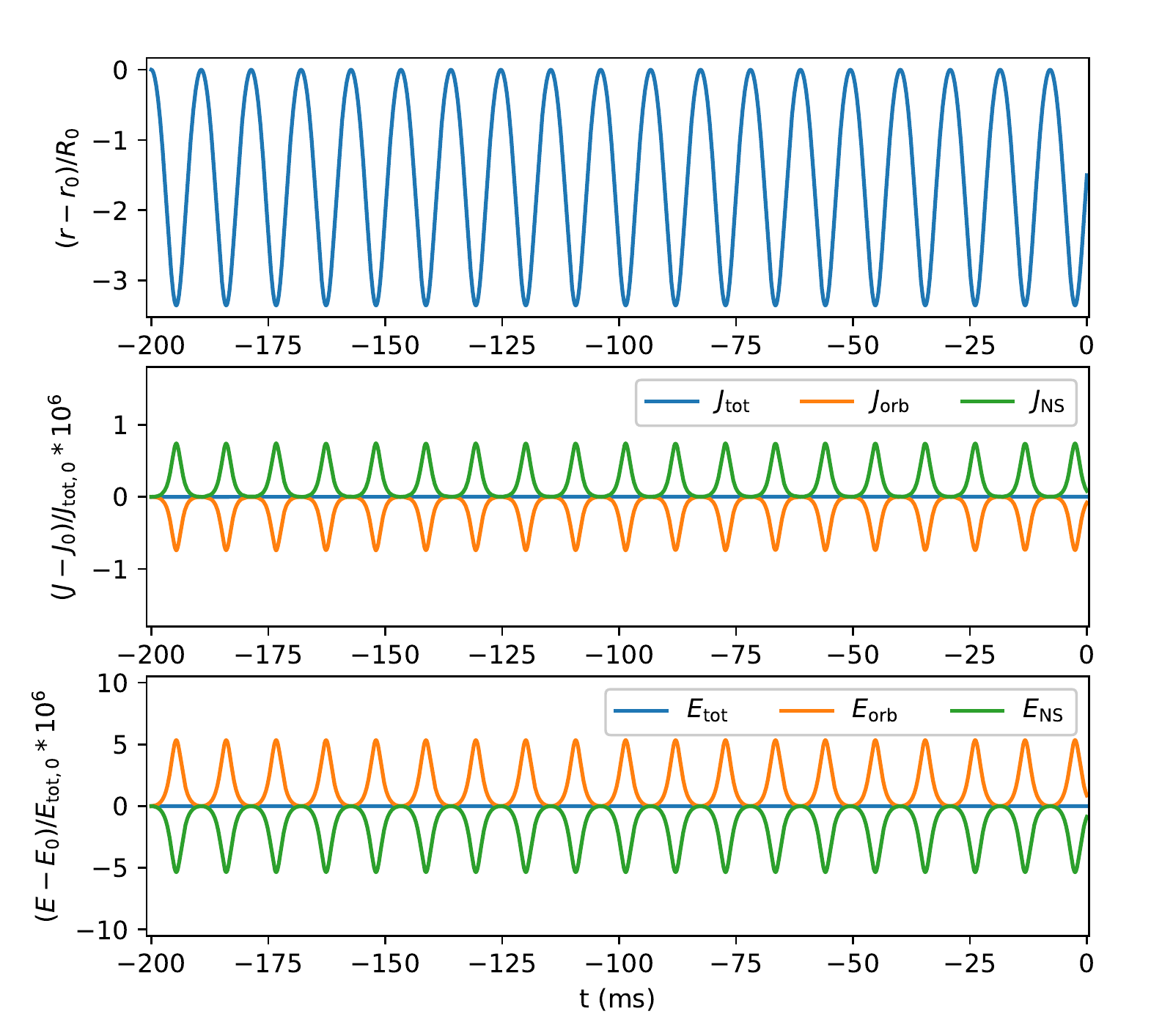}
	\caption{Conservation test of our integration for a pure
     Newtonian binary with the initial eccentricity $e_0=0.2$ and the initial (apocenter) separation $r_0=10.5R_0$. The three panels 
     show the evolution of binary separation ($r$), and various components of the energy and angular momentum of the system. The various quantities ($r,~J_{\rm orb},~J_{\rm NS},~E_{\rm orb},~E_{\rm NS}$) oscillate significantly with time, while $J_{\rm tot}$ and $E_{\rm tot}$ are conserved. Note that $J_{\rm tot,0}$ and $E_{\rm tot,0}$ are the initial total angular momentum and energy of the system.}
	\label{fig:conserve}
\end{figure}


\begin{figure*}[hbt!]
	\includegraphics[width=\textwidth]{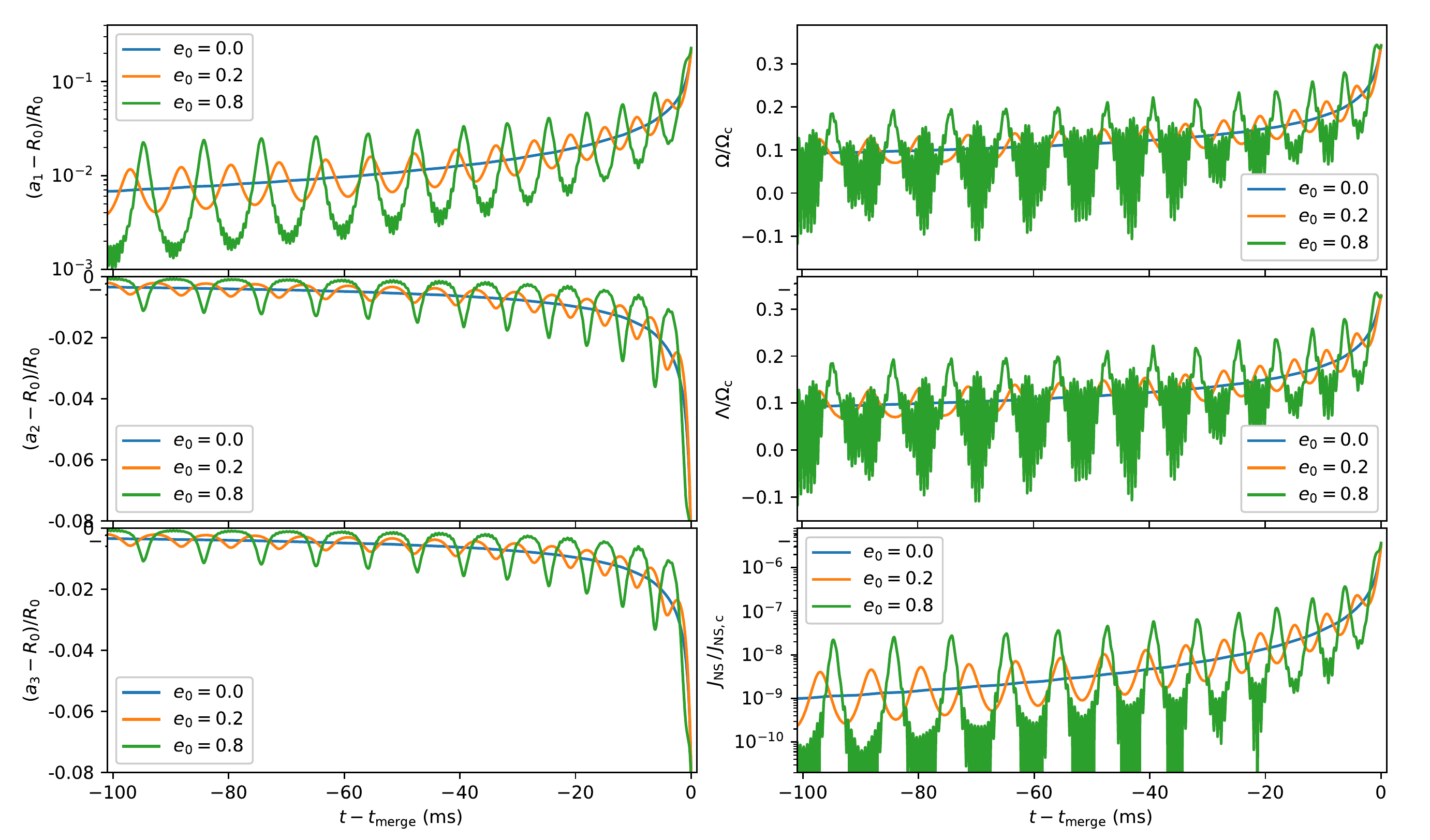}
	\caption{The last 100 ms dynamical evolution of the ellipsoid radii ($a_1,~a_2$, and $a_3$) and the NS angular frequency of the ellipsoidal figure ($\Omega$) and internal circulation ($\Lambda$), angular momentum ($J_{\rm NS}$) for a NS-BH inspiral with $e_0=0.0,~0.2,~0.8$. Note that $\Omega$ and $\Lambda$ are scaled by $\Omega_{\rm c}= \sqrt{\pi G\rho_0}$, and
	$J_{\rm NS}$ is scaled by the critical value $J_{\rm NS,c}= (2/5)\kappa_n MR_0^2\sqrt{\pi G\rho_0}$.}
	\label{fig:5}
\end{figure*}

\begin{figure*}[hbt!]
\includegraphics[width=\textwidth]{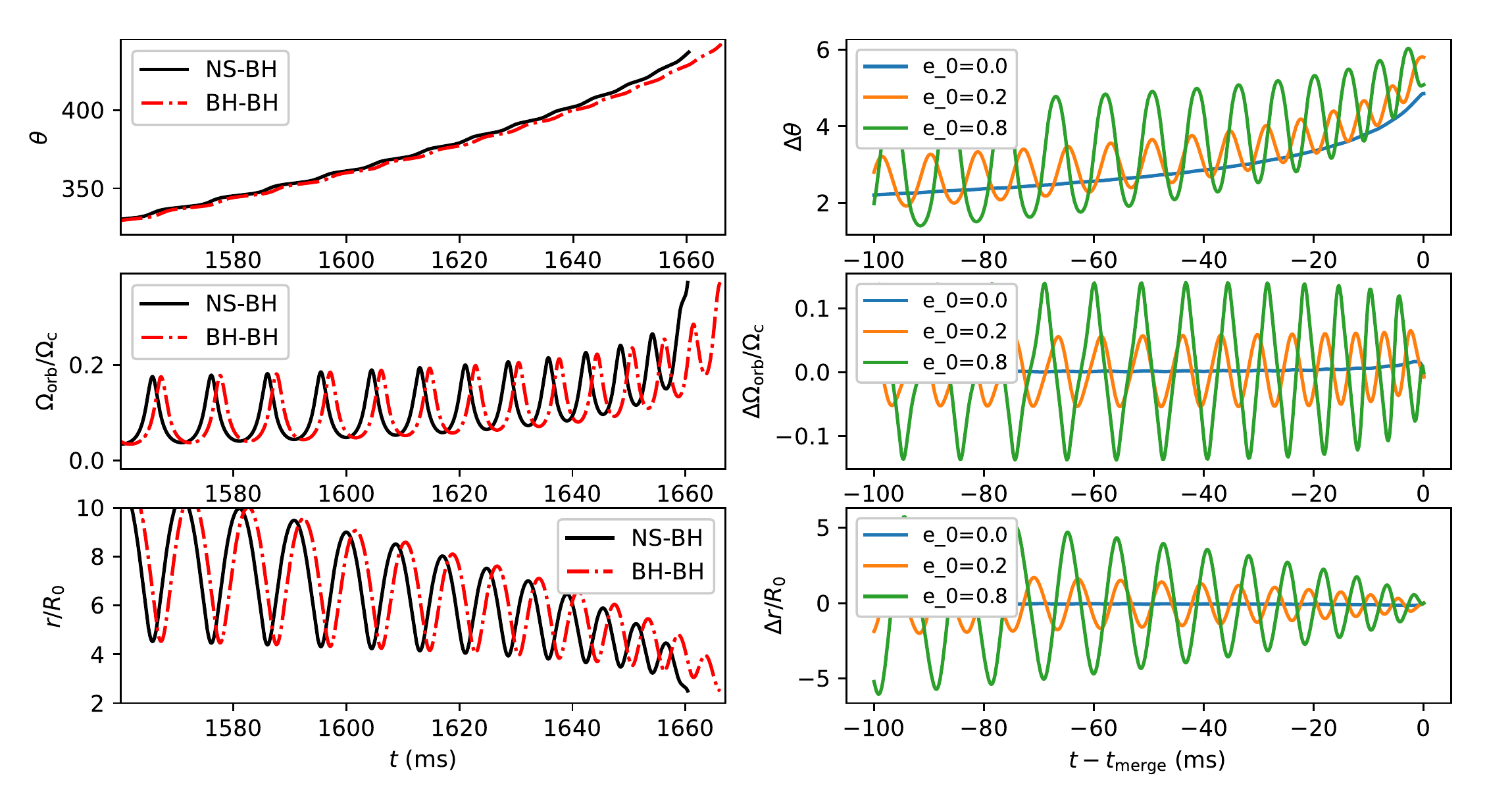}
\caption{Left panels: The last 100 ms of the dynamical evolution of
 the orbital phase ($\theta$), angular velocity ($\Omega_{\rm orb}$)
 and separation ($r$) for the BH-BH insprial (with no tidal effect) and NS-BH inspiral with $e_0=0.8$. Right panels: The shifts of orbital parameters between
 the BH-BH and NS-BH cases for $e_0=0.0,~0.2,~0.8$,
 where $\Delta \theta=\theta^{\rm BH-BH}-\theta^{\rm NS-BH}$, $\Delta
 \Omega_{\rm orb}=\Omega_{\rm orb}^{\rm BH-BH}-\Omega_{\rm orb}^{\rm
  NS-BH}$ (scaled by $\Omega_{\rm c}= \sqrt{\pi G\rho_0}$) and $\Delta r=r^{\rm BH-BH}-r^{\rm NS-BH}$. In all cases, the binary has initial pericenter distance $r_{p,0}=7R_0$, and starts at the apocenter with separation $r_0=r_{p,0} (1+e_0)/(1-e_0)$.}
	\label{fig:3}
\end{figure*}

The evolution equations (\ref{eq:a1})-(\ref{eq:theta}) can be stiff
because of the small quantity ($a_2/a_1-a_1/a_2$) that appears in the
denominators of Eqs.~(\ref{eq:omega}) and (\ref{eq:lambda}). To check
the accuracy of our integration, we first turn off all PN terms and
consider a pure Newtonian binary system by setting $A_H=B_H=A_{5/2}=B_{5/2}=0$. Fig. \ref{fig:conserve} shows the evolution
of the separation and various components of the angular momentum and
energy of the system for a binary with $e_0=0.2$. 
The angular momentum of the NS ($J_{\rm NS}$), the orbital angular
momentum ($J_{\rm orb}$), the energy of NS ($E_{\rm NS}$) and the orbital energy ($E_{\rm orb}$), as well as the total angular momentum and energy of the system ($J_{\rm tot}=J_{\rm NS}+J_{\rm orb}$, $E_{\rm tot}=E_{\rm NS}+E_{\rm orb}$) are expressed as Eqs. (2.4)-(2.9), (2.12) and (2.15) in Ref.~ \cite{Lai1995}. 
We see that our integration conserves energy and angular momentum to high precision
(better than $10^{-6}$).

We then solve the complete hybrid equations (\ref{eq:a1})-(\ref{eq:theta}), including 
the tidal effects and PN terms, for NS-BH inspirals with different initial eccentricities
(all with the same initial $r_{p,0}=7R_0$). 
Fig.~\ref{fig:5} shows the evolution of the ellipsoid radii, and the angular frequency of the ellipsoidal figure ($\Omega$), internal circulation ($\Lambda$), and the angular momentum of the NS. 
The tidal effects result in the elongation of $a_1$, but the contraction of $a_2$ and $a_3$. It also leads to significant oscillating behaviour of $\Omega$ and $\Lambda$, and a slight increase of the NS spin angular momentum as the orbit decays. 
The changes of these parameters grow as the orbit decays. For better visualisation, we focus on the last $100$ ms of the evolution in the figures, where these parameters change significantly.
For zero initial eccentricity, these parameters change gradually, while for larger initial eccentricities, their changes oscillate significantly near each pericenter passage, and strongly influenced by the f-modes. 



\begin{figure*}[hbt!]
\includegraphics[width=0.8\textwidth]{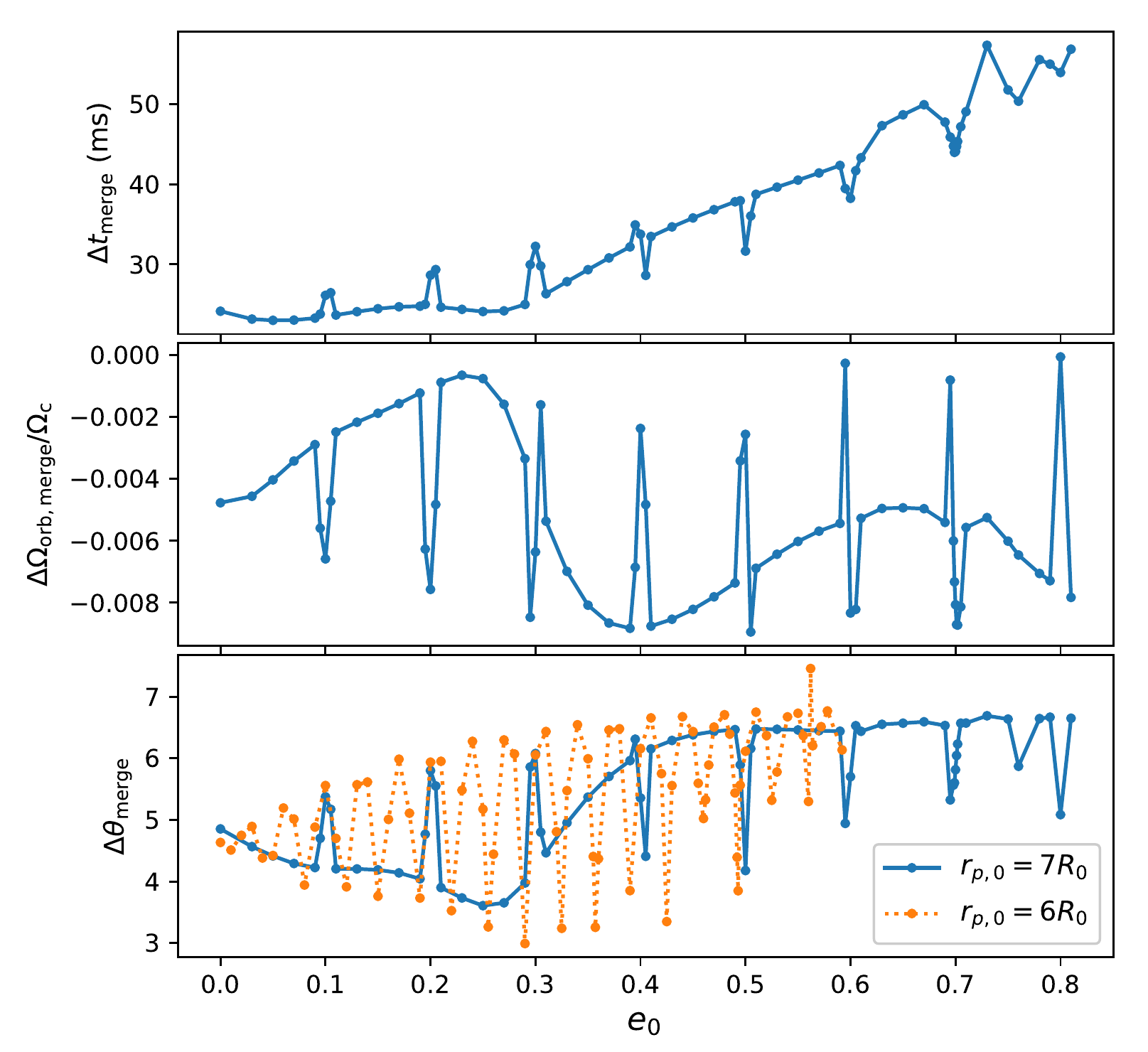}
\caption{The shifts of orbital parameters at the merger time between the BH-BH and NS-BH cases vs the initial eccentricity. Note that $\Delta \Omega_{\rm orb, merge}$ is scaled by $\Omega_{\rm c}= \sqrt{\pi G\rho_0}$. The initial pericenter distance is $r_{p,0}=7R_0$ for blue curves, and $r_{p,0}=6R_0$ for the orange dotted line in the bottom panel.}
	\label{fig:4}
\end{figure*}

To quantitatively examine the tidal effect on the orbital evolution, we compare the time evolution of the orbital phase ($\theta$), angular velocity ($\Omega_{\rm orb}$) and binary separation ($r$) between the NS-BH inspiral and BH-BH inspiral (with no tidal effect) in the left panels of Fig. \ref{fig:3}.
It can be seen that the tidal effects accelerate the merger, and influence the
evolution of the orbital parameters prior to the merger. We define the shifts of these parameters between the BH-BH and NS-BH cases as $\Delta x=x^{\rm BH-BH}-x^{\rm NS-BH}$ for $x=\theta,~\Omega_{\rm orb},~r,t$ to quantify these effects.
We show the evolution of the shifts for $e_0=0.0~,0.2,~0.8$ in the
right panels of Fig. \ref{fig:3}. For the circular case, the shifts
change gradually. For eccentric orbits, the shifts oscillate, and the
amplitudes of oscillation increase with the initial eccentricity. For
the orbital phase, we find that the tidal effects will always lead to
a positive shift of orbital phase ($\Delta\theta>0$).

Fig.~\ref{fig:4} shows the net shift at the merger (defined as $\Delta
x_{\rm merge}=x_{\rm merge}^{\rm BH-BH}-x_{\rm merge}^{\rm NS-BH}$)
as a function of the initial eccentricity. We see that the
tidal effects lead to faster mergers with $\Delta t_{\rm merge}\in(22,58)$ ms,
and higher final orbital frequencies with
$\Delta \Omega_{\rm orb, merge}\in(-86,0)$ rad/s. The cumulative orbital phase shifts at the merger are $\Delta \theta_{\rm merge}\in(3,7)$. 
Especially in some cases with $e_0\sim0.5-0.8$, the phase shift is $>2\pi$, meaning that the NS-BH case takes more than one orbit less than the BH-BH case to merge. 


\begin{figure}[hbt!]
\includegraphics[width=0.5\textwidth]{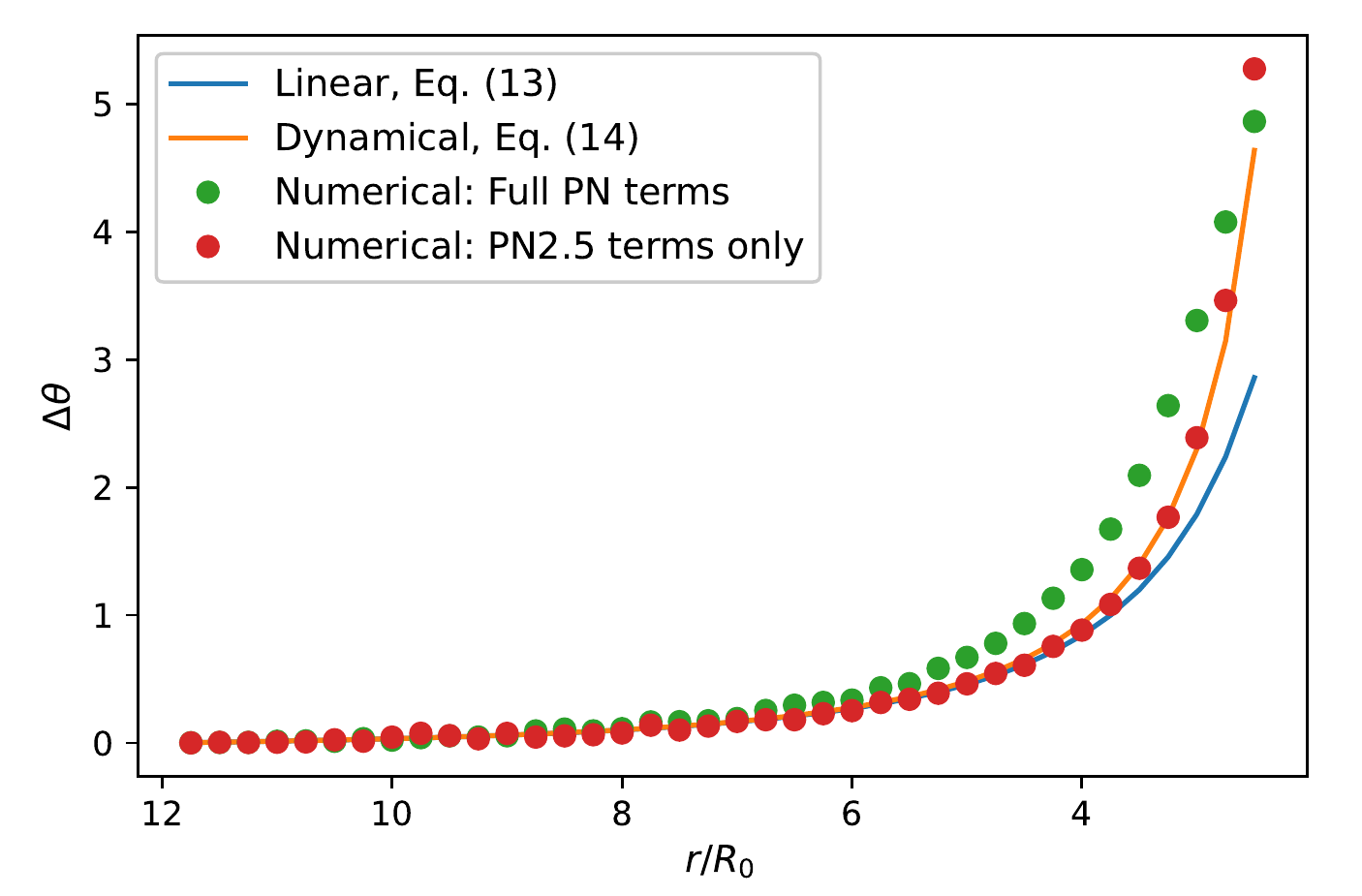}
\caption{The cumulative phase shift between BH-BH and NS-BH inspirals for circular orbits. The binary starts at the initial separation $r_0=12R_0$ and $e_0=0$.
The linear analytical results are given by Eq. (\ref{eq:dtheta_linear}) and (\ref{eq:dtheta_resonant}), respectively.
The ``numerical PN2.5 terms only'' result is obtained by setting $A_H=B_H=0$, but keeping $A_{5/2}$ and $B_{5/2}$ in Eqs. (\ref{eq:r})-(\ref{eq:theta}). The full numerical PN result is obtained by integrating Eqs. (\ref{eq:a1})-(\ref{eq:theta}) including all all PN terms. 
}
	\label{fig:dtheta}
\end{figure}

To contrast with previous works, we first consider the case of a circular orbit ($e_0=0$). The cumulative orbital phase shift in the linear tidal regime is given by (see Eq. (66) of Ref.\citep{Lai1994a}),
\begin{equation}
    \Delta \theta_{\rm Linear}^{\rm full}={1\over 16}k_2\left( {39\over4}+{M_{t}\over M'}\right){R_0^{5}\over M^2M_t^{1/2}}(r_f^{-5/2}-r_i^{-5/2}), \label{eq:dtheta_full}
\end{equation}
where $k_2=3\kappa_nq_n/2$ is the tidal Love number, and $r_i$ and $r_f$ are initial and final orbit separation. 
This phase shift is the same as the widely used expression given by Ref. \cite{Flanagan2008} after translating the binary separation $r$ into the orbital frequency (see Eqs.~(25) and (28) in Ref. \cite{Vick2019}). 
However, since our dynamical equations (1)-(7) do not include GW emission from the tidally deformed NS, the corresponding linear orbital phase shift is
\begin{equation}
    \Delta \theta_{\rm Linear}={39\over 64}k_2{R_0^{5}\over M^2M_t^{1/2}}(r_f^{-5/2}-r_i^{-5/2}). \label{eq:dtheta_linear}
\end{equation}
Comparing Eq. (\ref{eq:dtheta_full}) with Eq. (\ref{eq:dtheta_linear}), we see that the enhanced GW emission due to tidally deformed NS introduces only 9\% (for $M'\gg M$) or 17\% (for $M'=M$) correction to the tide-induced orbital phase shift.
The derivation of the above two equations assumes that the tidal forcing frequency ($2\Omega_{\rm orb}$) is much less than the intrinsic f-mode frequency ($\omega_f$) of the NS. 
Taking into account the dynamical response of the f-mode, the above equation becomes \cite{Xu2017},
\begin{equation}
    \Delta \theta_{\rm Linear}^{\rm dyn}=\Delta \theta_{\rm Linear}{1\over 1-4\Omega_{\rm orb}^2/\omega_{\rm f}^2},\label{eq:dtheta_resonant}
\end{equation}
where $\omega_{\rm f}=4\sqrt{\pi G\bar\rho/(15q_n)}$ (see Eq. (3.30) in Ref. \cite{Lai1994}). 
We compare our numerical results with these analytical results for circular orbits in Fig. \ref{fig:dtheta}. We study two cases: one with full PN terms as in Eqs. (\ref{eq:r}-\ref{eq:theta}), and the other with PN2.5 terms only (i.e. setting $A_H=B_H=0$, while keeping $A_{5/2}$ and $B_{5/2}$). 
The latter case is considered because of the fact that the analytical results (Eqs. \ref{eq:dtheta_full}-\ref{eq:dtheta_resonant}) are obtained without the non-dissipative first and second-order PN corrections.
We see that our numerical ``PN2.5 terms only'' result significantly deviates from $\Delta \theta_{\rm Linear}$ at small radii, but is consistent with $\Delta \theta_{\rm Linear}^{\rm dyn}$ except at $r\lesssim3R_0$. The ``PN2.5 terms only'' result also deviates from the ``Full PN terms" result at $r\lesssim7R_0$, with a deviation amount $\Delta \theta_{\rm Full~PN}- \Delta \theta_{\rm PN2.5}\lesssim1$. 
The nonlinear effect becomes important only very close to merger ($r\lesssim 3R_0$). Overall, Fig. \ref{fig:dtheta} indicates that the tide-induced phase shift is strongly influenced by the dynamical effect and by the non-dissipative PN effect on the orbit.

Looking back at Fig. \ref{fig:4}, we see that $\Delta \theta_{\rm merge}$ depends on $e_0$ in a non-monotonic way. 
The reason is that at the merger ($r=2.5R_0$), the orbit can still be eccentric, and the pericenter distance of the final orbit can be rather different between the NS-BH and BH-BH cases. As a result, the phase shift and orbit frequency shift would depend on the phase at the merger, i.e. whether the binary reaches $r=2.5R_0$ near the pericenter or apocenter. 
The behavior of $\Delta \theta_{\rm merge}-e_0$ for the $r_{p,0} =6R_0$ case \footnote{Note that for $r_{p,0}=6R_0$, we only show the result with $e_0<0.6$. The reason is that for large eccentricities with $r_{p,0}=6R_0$, the initial condition is not sufficiently accurate (as the initial condition is set by the equilibrium state), which makes the integration inaccurate.} is largely similar to the result obtained in 
Ref. \cite{Vick2019} using the linear tidal theory, where $\Delta \theta_{\rm merge}<0$ is observed for some eccentricities.
In our example depicted in Fig. \ref{fig:4}, $\Delta \theta_{\rm merge}>0$ is found for all values of $e_0$. 
This can be attributed to the fact that we consider the NS with $n=0.5$ in this paper, whereas Ref. \cite{Vick2019} adopts $n=1$.
The difference in the results between the case of $r_{p,0}=7R_0$ and that of $r_{p,0}=6R_0$ arises from the fact that the more compact initial orbits can retain a higher eccentricity near merger, giving rise to a stronger dependence of $\Delta\theta$ on the orbital phase at merger.

\begin{figure*}[ht]
\includegraphics[width=0.9\textwidth]{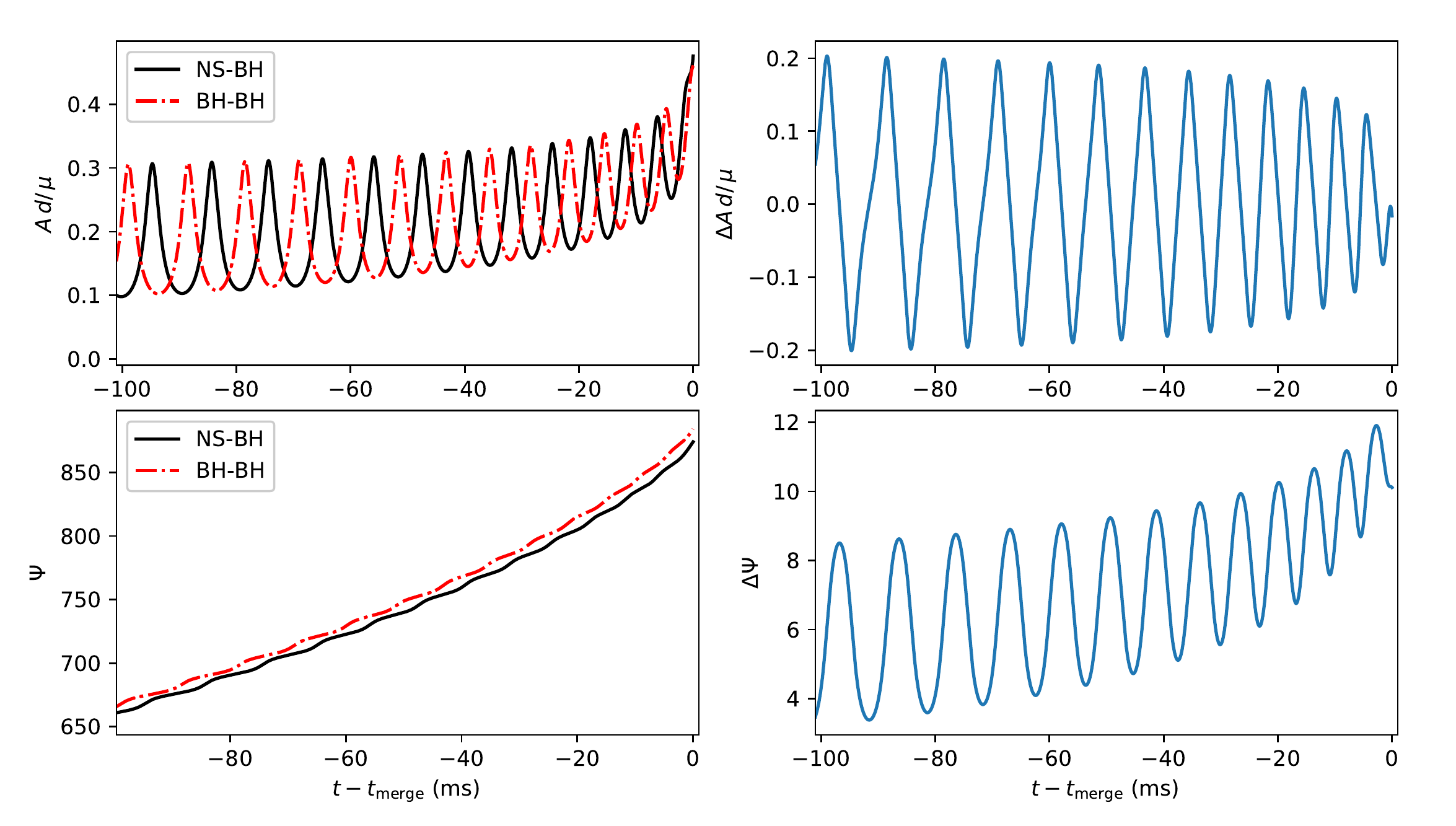}
\caption{The GW waveforms for inspirals with $e_0=0.8$ and the initial $r_{p,0}=7R_0$. 
Left panels: The evolution of GW amplitude and phase for NS-BH and BH-BH inspirals.
Right panels: The evolution of the amplitude difference $\Delta A$ and cumulative phase shift $\Delta \Psi$ between the BH-BH and NS-BH cases.}
	\label{fig:6}
\end{figure*}
\section{Effects on gravitational waves\label{GW}}

The leading-order gravitational wave forms of the two polarisation modes are given by \cite[e.g.][]{Vick2019}
\begin{eqnarray}
h_+ &=& \frac{(1+\cos^2{\Theta}) \mu}{d}(\dot{r}^2 \cos{2 \theta} + r \ddot{r} \cos{2 \theta} - 4 r \dot{r} \Omega_{\rm orb} \sin{2 \theta} \nonumber\\
& & -2r^2\Omega_{\rm orb}^2 \cos{2 \theta}-r^2\ddot{\theta} \sin{2 \theta}), \label{eq:hplus}\\
h_\times &=& \frac{2\cos{\Theta}\mu}{d}(\dot{r}^2 \sin{2\theta} + r \ddot{r} \sin {2\theta} +4 r\dot{r}\Omega_{\rm orb} \cos{2\theta}  \nonumber\\
& &- 2 r^2 \Omega_{\rm orb}^2\sin{2 \theta} + r^2 \ddot{\theta} \cos{ 2 \theta}). \label{eq:hcross}
\end{eqnarray}
where $d$ is the distance to the source, $\Theta$ is the angle
between the line of sight and the orbit normal direction
(the $z$-axis), and we have neglected the contribution from the quadruple moment of the NS.
For simplicity, we consider $\Theta=0$. The GW amplitude
($A$) and phase ($\Psi$) are given by 
\begin{equation}
A e^{-\text{i} \Psi(t)} = h_+ - \text{i} h_\times. \label{eq:Adef}
\end{equation}

Fig. \ref{fig:6} shows the time evolution of the amplitudes and
phases of the BH-BH and NS-BH inspirals in the left panels; in the
right panels, we show the difference between the two cases, i.e.
$\Delta x=x^{\rm BH-BH}-x^{\rm NS-BH}$ for $x=A,~\Psi$. We see 
that the tidal effects can significantly modify both the amplitude and the phase evolution. For the case of $e_0=0.8$, the amplitude difference oscillates considerably with $|\Delta A| /A$ reaching as large as $40\%$. 
The cumulative phase shift also oscillates significantly just like the orbital phase (see Fig. \ref{fig:3}, the top right panel), and the phase shift at the merger reaches $\Delta \Psi\sim12$.

\begin{figure*}[hbt!]
	\includegraphics[width=0.6\textwidth]{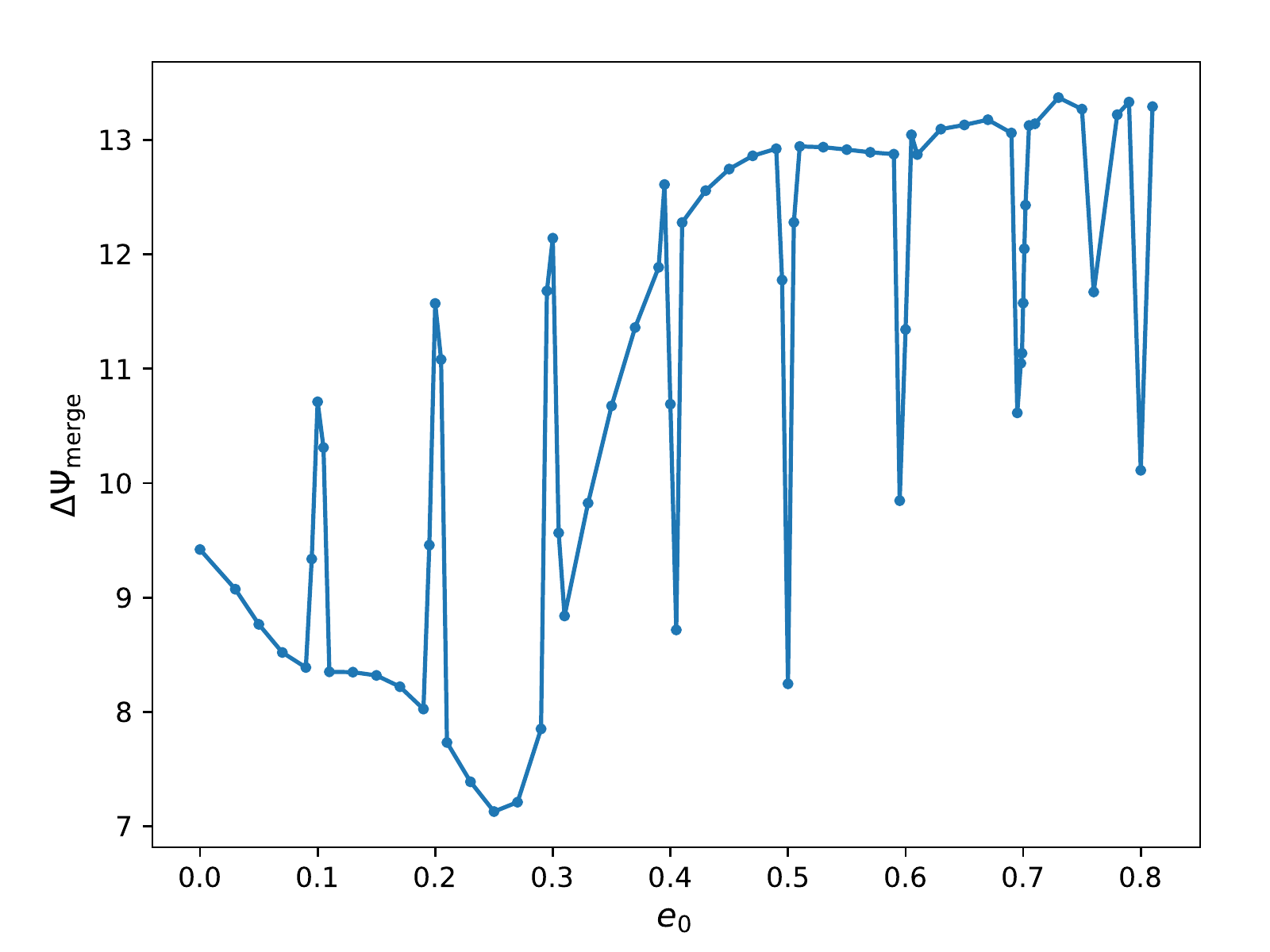}
	\caption{The cumulative phase shift $\Delta \Psi_{\rm merge}$ between the BH-BH and NS-BH inspirals as a function of the initial eccentricity for binaries with initial $r_{p,0}=7R_0$.}
	\label{fig:7}
\end{figure*}

Fig.~\ref{fig:7} shows the cumulative phase shift ($\Delta \Psi_{\rm merge}$) at the merger. 
We see that the tidal effects in the NS-BH binary lead to a significant phase shift $\Delta \Psi_{\rm merge}\in(7.1,13.4)$. 
Note for the circular case, the GW phase shift $\Delta \Psi_{\rm merge}=2 \Delta \theta_{\rm merge}$. 
The oscillating behaviour in the $\Delta \Psi_{\rm merge}-e_0$ plot is similar to that in the $\Delta\theta_{\rm merge}-e_0$ plot (see Fig. \ref{fig:4}), and is caused by the strong dependene of $\Delta\theta$ on the orbital phase at the merger if the binary is not fully circularised at $r=2.5R_0$. 


\section{Conclusion\label{Conclusion}}

In this paper, we have studied the tidal effects in eccentric inspiralling NS binaries, where the NS is modelled as a Newtonian compressible ellipsoid (which can deform non-linearly in response to tidal forcing) and the PN terms (up to the 2.5-PN order) are incorporated for the orbital evolution. 
Our treatment in this paper complements the linear mode approach in our recent work \citep{Vick2019} and other related works (see Section 1). 
We find that the tidal effects can accelerate the inspiral process, and induce orbital frequency and phase shifts. 
It can also lead to the elongation of $a_1$, but the contraction of $a_2$ and $a_3$. 
The change of ellipsoid radii at the merger can reach $\sim10\%$. 
For eccentric orbits, these shifts oscillate significantly around pericenter passages, and the amplitudes of the oscillation increase with eccentricities. 
In contrast, for circular inspirals, the phase shift evolves gradually as the orbit decays, and our calculation reproduces previous linear result at large binary separations but also indicates that the dynamical tidal response of the NS is important at small separations. 
At the merger, the cumulative orbital phase shift reaches $\Delta \theta_{\rm merge}\in(3,7)$ for different values of initial eccentricities at the initial pericenter distances of 7$R_0$ and canonical NS parameters (mass $M=1.4M_\odot$ and radius $R_0=11.6$ km). 
In particular, for some cases with $e_0\sim0.5-0.8$, the phase shift can be larger than $2\pi$, implying that without the tidal effects, the binary will undergo at least one more orbit.

The tidal effect on the GW is also significant, as it can induce significant 
phase shift with $\Delta \Psi_{\rm merge}\in(7.1,13.4)$ at the merger, as shown in Fig. \ref{fig:7}. 
For most eccentric insprials, the phase shifts are much larger than that of circular inspirals. Overall, these results are consistent with the previous calculations based on linear models (e.g. \cite{Vick2019}).

As noted in Section 1, the event rate of eccentric NS binary mergers is highly uncertain. The results presented in this paper show that if such systems (with sufficiently large eccentricities) are detected, they would provide useful information on the NS equation of state through the enhanced tide-induced phase shift.
But this would require more accurate waveform templates for eccentric NS mergers, due to the non-monotonic dependence of $\Delta \Psi_{\rm merge}$ on $e_0$.

\begin{acknowledgments}
We thank the referee for valuable comments. JSW is supported by China Postdoctoral Science Foundation (Grant 2018M642000, 2019T120335).
\end{acknowledgments}

\bibliography{ref}

\end{document}